\documentclass[twocolumn]{aastex631}

\graphicspath{{./}{figures/}}
\usepackage{graphicx}
\usepackage{gensymb}
\usepackage{footnote}

\begin{document}

\title{Probing the Hot Gaseous Halo of the Low-mass Disk Galaxy NGC\,7793 with eROSITA and Chandra}
\author[0000-0002-7875-9733]{Lin He}
\email{helin@smail.nju.edu.cn}
\affiliation{School of Astronomy and Space Science, Nanjing University, Nanjing 210023, China}
\affiliation{Key Laboratory of Modern Astronomy and Astrophysics, Nanjing University, Nanjing 210023, China}

\author[0000-0003-0355-6437]{Zhiyuan Li}
\email{lizy@nju.edu.cn}
\affiliation{School of Astronomy and Space Science, Nanjing University, Nanjing 210023, China}
\affiliation{Key Laboratory of Modern Astronomy and Astrophysics, Nanjing University, Nanjing 210023, China}
\affiliation{Institute of Science and Technology for Deep Space Exploration, Suzhou Campus, Nanjing University, Suzhou 215163, China}

\author[0000-0001-9062-8309]{Meicun Hou}
\affiliation{Institute of Science and Technology for Deep Space Exploration, Suzhou Campus, Nanjing University, Suzhou 215163, China}

\author[0000-0001-9953-0359]{Min Du}
\affiliation{Department of Astronomy, Xiamen University, Xiamen, Fujian 361005, China}

\author[0000-0002-2853-3808]{Taotao Fang}
\affiliation{Department of Astronomy, Xiamen University, Xiamen, Fujian 361005, China}

\author[0000-0002-6324-5772]{Wei Cui}
\affiliation{Department of Astronomy, Tsinghua University, Beijing 100084, China}

\begin{abstract}
Galaxy formation models predict that local galaxies are surrounded by hot X-ray-emitting halos, which are technically difficult to detect due to their extended and low surface brightness nature. Previous X-ray studies have mostly focused on disk galaxies more massive than the Milky Way, with essentially no consensus on the halo X-ray properties at the lower mass end. We utilize the early-released eROSITA and archival {\it Chandra} observations to analyze the diffuse X-ray emission of NGC\,7793, a nearby spiral galaxy with an estimated stellar mass of only $\rm 3.2\times 10^9~M_{\odot}$. We find evidence for extraplanar hot gas emission from both the radial and vertical soft X-ray intensity profiles, which spreads up to a galactocentric distance of $\sim 6$ kpc, nearly $30\%$ more extended than its stellar disk. Analysis of the eROSITA spectra indicates that the hot gas can be characterized by a temperature of $0.18^{+0.02}_{-0.03}$ keV, with 0.5--2 keV unabsorbed luminosity of $\rm 1.3\times 10^{38}~erg~s^{-1}$. We compare our results with the IllustrisTNG simulations and find overall consistence on the disk scale, whereas excessive emission at large radii is predicted by TNG50. This work provides the latest detection of hot corona around a low-mass galaxy, putting new constrains on state-of-the-art cosmological simulations. We also verify the detectability of hot circumgalactic medium around even low-mass spirals with future high-resolution X-ray spectrometer such as the Hot Universe Baryon Surveyor.

\end{abstract}

\section{Introduction} \label{sec:intro}
It is a long-standing prediction of galaxy formation models that a rarefied, volume-filling hot gas corona exists around most present-day (disk) galaxies \citep{White_78}, giving rise to the soft X-ray emission with a characteristic temperature of $10^{6-7}$ K. Two prevailing scenarios have been proposed for its origin: In an external way, the intergalactic medium (IGM) gas confines to the galaxy and gets gravitationally heated, leaving a hot halo surrounding the disk when the gas cooling time exceeds the free-fall timescale \citep{White_78,White_91}. In an internal channel, galactic feedback processes, such as stellar winds and supernovae (SNe), thermalize and metal-enrich the ambient medium, further entrain them to the halo via the driving of outflows/winds \citep{Mathews_71,Chevalier_85}. 
In the meantime, feedback from an active galactic nucleus (AGN) can also impact the hot gaseous halo by driving vertical outflows which can transport the energy, momentum and metals \citep{Choi_20}.
As one would expect, this X-ray-emitting hot corona (also known as hot circumgalatic medium [CGM]) is a very complicated reservoir mixing the accreted IGM, the star-formation-enriched outflowing gas and the recycling flows, and it potentially occupies a significant fraction of the cosmic missing baryons.

In spite of their fundamental importance, the detection and characterization of galactic hot gas halos is challenging, owing to a low gas density. The thermal Sunyaev-{Zel\'dovich} (SZ) effect and soft X-ray emission are two routinely used tools to explore this hot phase CGM. The former, aiming to measure the distortion of the Cosmic Microwave Background (CMB) power spectrum arising from the inverse Compton scattering of CMB photons by free electrons, is generally sensitive to massive clusters/groups and the measurements of CGM have used ensembles of galaxies splitted into different mass bins to achieve a reliable detection \citep{Bregman_18,Bregman_22,Planck_2013,Greco_15}. On the other hand, while the hot gas halo emits in soft X-rays, the signals can be very weak and often overwhelmed by galactic stellar sources. Efforts have been made to detect the diffuse X-ray emission in nearby spiral galaxies \citep{LWH_07, Li_07, Dai_12, AB_11, Bogdan_13b, Bogdan_13a, Walker_15, Li_16, Li_17}, relying on various X-ray missions (ROSAT, {\it Chandra} and XMM-Newton), and it has led to the consensus that hot gas halos are ubiquitous in massive spirals, with their X-ray luminosity scaled with the star formation rate (SFR) \citep{Li_13b,Li_14}. 
Moreover, dedicated studies for a few giant spirals (e.g. NGC 1961, \citealt{Anderson_16} ; NGC 6753, \citealt{Bogdan_17}) find a negative gradient of the temperature and a rather flat, sub-solar metallicity profile for the galactic hot gas, hinting at the predominant role of IGM accretion as its origin.

However, there is less consensus on the physical properties (e.g. mass, temperature, metallicity) and the spatial distribution of hot gas in and around less massive $L_*$ galaxies (especially for $M_* < M_{\rm MW}$). In theory, the virialized halo gas will have a temperature of $T_{\rm v}=\frac{2}{3}\frac{GM_{\rm v}}{r_{\rm v}}\frac{\mu m_{\rm H}}{k_{\rm B}}$ \citep{Benson_2010}, where $M_{\rm v}$ and $\rm r_{\rm v}$ are the galaxy's virial mass and radius, $G$ and $k_{\rm B}$ are the gravitaional and Boltzmann constant, $\mu$ is the mean molecular mass, and $m_{\rm H}$ is the atomic mass unit. Thus galaxies at the low-mass end are predicted to have halos with lower temperature, less emissivity, and are more difficult to detect. 
Alternatively, stellar feedback, in the form of galactic outflows/winds, is thought to play a key role in the gas replenishment of the hot corona surrounding less massive galaxies. Diffuse X-ray emission emerging from the hot coronal gas has been detected in both starburst dwarf galaxies \citep{Ott_05} and archetypical starburst disk galaxies (e.g., M82, \citealp{Lehnert_1999}; NGC\,253 \citealp{Strickland_2002}), with its spatial distribution elongated along the direction of the maximum pressure gradient. Additional examples can be found in a systematic study of galactic coronae around nearby highly inclined disk galaxies conducted by \citet{Li_13a}, which covers a broad range of SFR and stellar masses. It would be valuable to investigate if the feedback-related hot coronae are ubiquitous from actively star-forming dwarf galaxies to normally star-forming $L_*$ galaxies, and if they share a common origin.

A major limitation is that current sample of galactic hot gaseous halo has modest size and is biased to massive disk galaxies in the local universe. Specifically, there is still an unsatisfactory lack of individual detections in low-mass galaxies (e.g., non-detections in \citealt{Bothun_1994, Bomans_2001, Mulchaey_2010, Hou_21, Hou_2024}). 
Stacked measurements of this diffuse and faint component have been obtained thanks to the wide-field X-ray survey such as the ROSAT All-Sky Survey (RASS), the eROSITA Final Equatorial Depth Survey (eFEDS) and its all-sky survey (eRASS, \citealt{Bogdan_15,Anderson_13,Chadayammuri_22,Zhang_2024a,Zhang_2024b,Zhang_2024c}). By accumulating signals from a large number of galaxies, the diffuse X-ray emission arising from the hot CGM can be captured and is found to spread out to hundreds of kiloparsec for galaxies at various mass bins. 
However, drawbacks of this method are obvious: (1) the resolution of stacked signals are generally poor ($\sim 10$ kpc for ROSAT stackings of Milky Way size galaxies at a median redshift of $z \sim 0.08$), and hence can not bring out the internal structure of CGM; (2) the removal of contribution from groups and clusters along the line of sight poses theoretical and technical challenges, hence the contamination is inevitable. For this reason, individual measurements are still valuable to help us better understand the galaxy formation theory and put tighter constraints on state-of-the-art cosmological simulations.

Launched in 2019, eROSITA (extended ROentgen Survey with an Imaging Telescope Array) was designed to perform sensitive wide-field sky surveys, with sensitivity enhanced more than thirtyfold compared with ROSAT \citep{Predehl_21}. This telescope contains seven identical Wolter-1 mirror modules, achieving an effective area (2400 $\rm cm^2$ at 1 keV) much larger than that of {\it Chandra}  and a moderate angular resolution ($\rm 15\arcsec$ on-axis) comparable to XMM-Newton. 
The large field of view of eROSITA ($\sim$ 1 degree in diameter) will enable us to depict the large-scale vicinity of nearby galaxies and track the elusive hot X-ray corona out to large radii, which is costly for {\it Chandra} and XMM-Newton. It holds promise to detect the diffuse X-ray emission from a large number of galaxies and to provide a statistically representative sample for galactic X-ray studies. 

NGC\,7793, with a stellar mass of $\rm 3.2\times 10^9~M_{\odot} $\citep{Bothwel_09} and an SFR of 0.52 $\rm M_{\odot}~yr^{-1}$ \citep{Calzetti_15}, is one of the brightest galaxies in the Sculptor Group, the nearest constellation of galaxies outside our Local Group. It is classified as an Sd galaxy with a very tiny bulge and chaotic spiral arms. This bulgeless charateristic, along with its moderate stellar mass, makes it controversial to fuel a MBH and contain large volumes of hot gas. Besides, this galaxy is famous for harboring a powerful microquasar \citep{Pakull_10}, which ejects collimated jets strong enough to inflate a fast-expanding bubble spanning $\sim$ 300 parsecs in diameter. In virtue of its proximity (at a distance of 3.91 Mpc, \citealt{Karachentsev_03}) and the peculiarity, it has a wealth of data including HST, Spitzer, GALEX, VLA, MUSE, XMM-Newton and $\it Chandra$, along with a planned JWST observation.
Deep observations have shown that its $\rm H\alpha$ and $\textsc{H\,I}$ emission easily extend to or beyond the optical radius $R_{25}$, both manifesting a truly declining rotational curve \citep{CP_90,Dicaire_08}. The diffuse ionized gas 
accounts for $\sim 40\%$ of the its $\rm H\alpha$ emission, and can be detected reaching the edge of the $\textsc{H\,I}$ disk, making the $\textsc{H\,II}$ disk of this galaxy one of the largest ever observed among the non-active late-type systems.
However, its galactic X-ray emission remains relatively underexplored, especially for the measurement of halo gas. \citet{Pannuti_11} conducted a {\it Chandra} ACIS observation to study the supernova remnants (SNRs) in NGC\,7793. In addition, they found that the galaxy's diffuse X-ray emission can be characterized by a temperature of 0.19--0.25 keV. A ROSAT PSPC observation \citep{RP_99} has detected the rather uniform unresolved emission in and around NGC\,7793, extending out to $\sim$ 4 kpc with a higher temperature of $kT \sim$ 1 keV. The discrepancy of the gas temperature may primarily lie in the much broader point spread function (PSF) wings of ROSAT PSPC, which led to the contamination of more spillover harder X-ray photons from the point sources. Moreover, due to the relatively poor sensitivity of ROSAT, the diffuse emission is also heavily contaminated by unresolved stellar sources, which is expected to be mitigated in the eROSITA era.

This paper is organized as follows: in Section \ref{sec:data prep}, we describe the observation and data preparation. The results and analysis, including the radial (vertical) perspective of the diffuse X-ray emission distribution and spectral analysis, are presented in Section \ref{sec:results}. We compare our results with TNG50 simulations in Section \ref{sec:TNG50} and discuss the implications in Section \ref{sec:discussion}. Our main conclusions are summarized in Section \ref{sec:summary}. Throughout this work, we assume a flat $\rm \Lambda CDM$ cosolomogy with $\Omega_m=0.3$, $\Omega_{\Lambda}=0.7$, and $H_0=70\ \rm km\ s^{-1}\ Mpc^{-1}$. All uncertainties are quoted at $1\sigma$ confidence level unless stated otherwise.

\section{Observation and Data Preparation}\label{sec:data prep}

\subsection{{\it eROSITA} }\label{subsec:eROSITA data}

NGC\,7793 was observed by eROSITA as part of the Calibration and Performance Verification (Cal-PV) program. A pointed observation (ObsID: 300011) was conducted on NGC\,7793 utilizing all 7 eROSITA telescope modules (TMs), with an effective exposure time of $\sim$ 51.2 ks and a covered field of diameter $\sim 1\deg$. 
The eROSITA data reduction was performed using the extended Science Analysis Software System (eSASS, \citealt{Brunner_22}) and the eROSITA Calibration Database (CALDB). 
As the TM5 and TM7 are reported to be heavily contaminated by light leak at low energies, we adopt an energy filter of PI $ > $ 400 to avoid this effect following previous eROSITA studies\footnote{We have compared the results when excluding and including the light-leak affected TM5 $\&$ TM7 and found no significant differences. We thus retain all the seven TMs but adopt a relatively higer lower energy bound, in order to better probe the putative faint diffuse emission. Similar treatments to the light leak contamination can also be found in \cite{Zhang_2024a} and \cite{Veronica_2024}}. 
We focus our analysis on three energy bands: the soft (0.4--1.0 keV), hard (1.0--2.3 keV) and full band (0.4--2.3 keV). We first generated the counts and vignetting-corrected exposure maps in each band using the {\it evtool} and {\it expmap} task, merging data derived from all 7 TMs. The argument gti=``FLAREGTI" was additionally adopted to filter out the potential soft proton flares. 
Next we created the corresponding instrumental background maps from the Filter Wheel Closed (FWC) data. To be consistent with the released calibrated event file, we specified flag=0xC000F000 to remove strongly vignetted corners of the square CCDs and pattern=15 to select single, double, triple, and quadruple patterns. The instrumental background for all the 7 TMs were merged, preparing for the following analysis. 

\citet{Liu_22} carried out source detection for 11 extra-galactic CalPV fields and presented a serendipitous catalog containing 9522 X-ray sources, among which 647 individuals are in the N7793 field. 
In view of its relatively large point spread function (PSF) wings, we masked the detected sources by circles of the 90\% enclosed energy radius (EER), adjusting to 1.5 times 90\% EER for some very bright point sources. 
All the counts, exposure and instrumental background maps were uniformly filtered off the detected X-ray sources for the following analysis. 
It is noteworthy that this procedure still cannot completely eliminate a few extended sources, including SNRs in the disk of NGC\,7793 and background galaxy groups or clusters, which may contribute to some excesses in the final intensity profiles (see details in Section \ref{subsec:rprofile} and Section \ref{subsec:vprofile}). 

\begin{deluxetable}{ccccc}
\tabletypesize{\scriptsize}
\tablecaption{X-ray Observations of NGC\,7793 \label{tab:obs}}
\tablewidth{0pt}
\tablehead{
\colhead{Observing Date} & \colhead{ObsID} & \colhead{Exposure} & \colhead{R.A.} & \colhead{Dec.}
}
\startdata
 &  & eROSITA &  &  \\
 \hline
2019-11-18 & 300011-1 & 53.8 & 23:57:51.00 & -32:37:27.0 \\
\hline
 &  & {\it Chandra} &  &  \\
 \hline
2003-09-06 & 3954 & 42.5 & 23:57:49.75 & -32:35:29.5 \\
2011-08-13 & 14231 & 58.5 & 23:57:59.90 & -32:33:20.9 \\
2011-12-25 & 13439 & 57.4 & 23:57:59.90 & -32:33:20.9 \\
2011-12-30 & 14378 & 24.7 & 23:57:59.90 & -32:33:20.9 \\
2020-06-04 & 23266 & 29.7 & 23:57:51.00 & -32:37:26.6 \\
\enddata
\tablecomments{The exposure time for eROSITA is the effective unvignetted exposure time averaged over the 7 TMs. The exposure times for {\it Chandra} have values after filtering the soft proton flares. All exposures are in units of ks. 
}
\end{deluxetable}

\subsection{{\it Chandra}}\label{subsec:Chandra data}

NGC\,7793 has five publicly available {\it Chandra} observations by October 2022.
All five observations had their aim-point on the S3 CCD of the Adanvanced CCD Imaging Spectrometer (ACIS), which was placed within $3\arcmin$ from the galactic center.
We included data from the S2, S3, I2 and I3 CCDs to maximize the useful field-of-view (FoV; covering a galactocentric radius up to $\sim 15\arcmin$) and the signal-to-noise ratio (S/N). The data reduction is performed using CIAO v4.13\footnote{\url{http://cxc.harvard.edu/ciao/}} and CALDB v4.9.5, following the standard procedures as described in \citet{Hou_21}. We produce the counts and exposure maps on the natal pixel scale of $0.492\arcsec$ in the 0.5--1, 1--2 and 0.5--2 keV band, from which the time intervals contaminated by soft proton flares were filtered out for each observation, resulting in a total effective exposure of $\sim$ 210 ks. We also generated the corresponding instrumental background maps from the stowed background files, after normalizing with the 10--12 keV count rate. The counts, instrumental background and exposure maps were then reprojected to a common tangential point, i.e., the optical center of the galaxy to produce the combined images.

\begin{figure*}
\centering
\includegraphics[width=1.\textwidth]{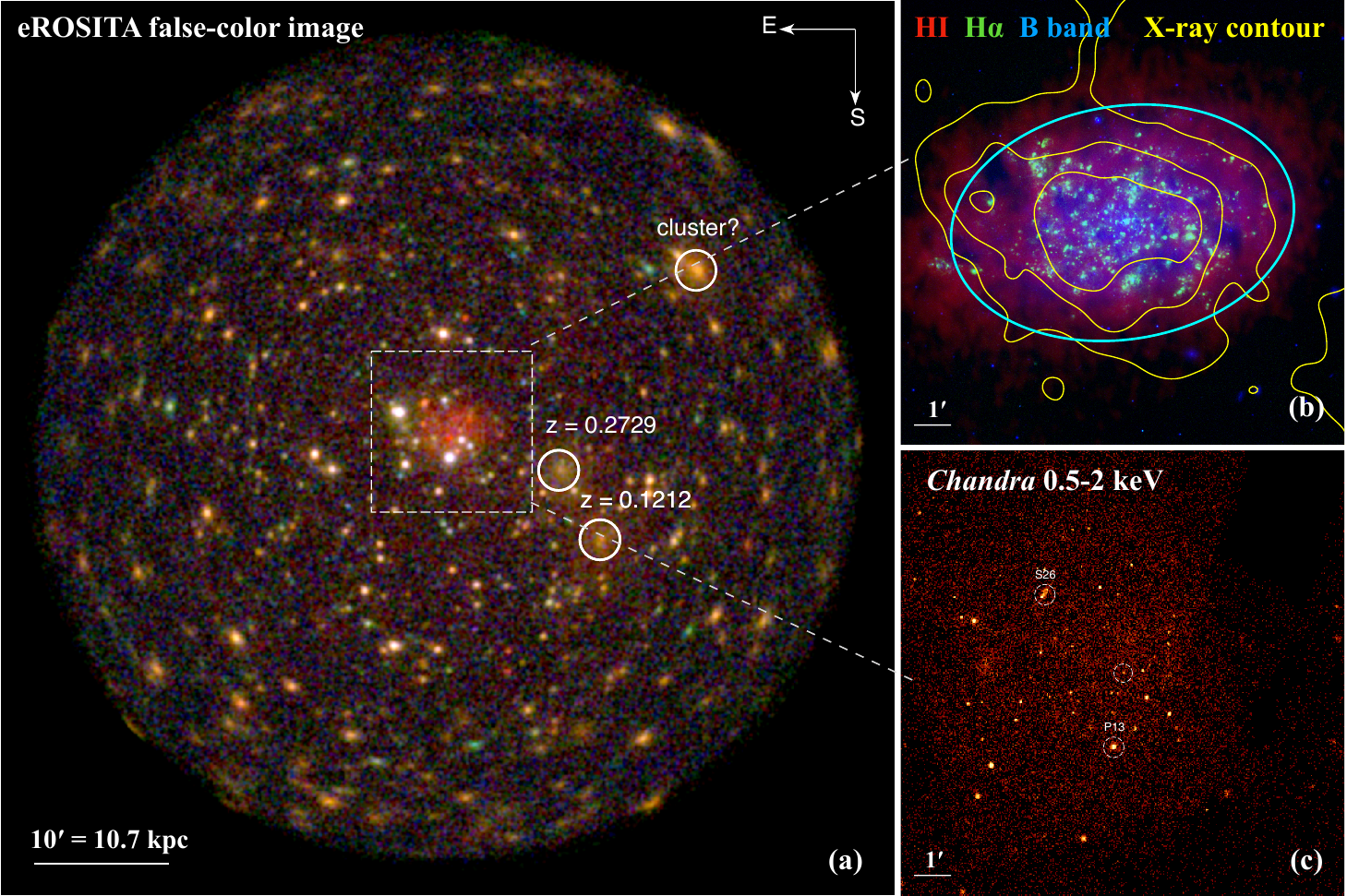}
\caption{{\it Left}: the eROSITA false-color image of the NGC\,7793 field. The red channel corresponds to 0.4--1 keV, green to 1--2.3 keV, and blue to 2.3--5 keV. The white dashed box encloses the central $12\arcmin \times 12\arcmin$ region. Two background clusters in southwest are circled and deliberately omitted in the analysis. {\it Upper right}: The composite image of $\textsc{H\,i}$ (red), $\rm H\alpha$ (green) and B-band (blue), with the 0.4--2.3 keV diffuse X-ray emission in  yellow contours (derived from eROSITA flux map with the location of removed point sources refilled by sampling their surrounding background), at intensity levels of 3.5, 7.0, 10.5 $\times 10^{-4}\rm~ cts~s^{-1}~arcmin^{-2}$. The $D_{25}$ ellipse in cyan has a diameter of {\bf $9.3\arcmin \times 6.3\arcmin$}. 
{\it Lower right}: the combined Chandra counts map in the energy band of 0.5--2.0 keV, zoomed-in as the composite image. The ULX P13 and the nebula S26 are marked with white dashed circles. Also marked by a dash circle is the X-ray nucleus, which, despite being faint, is revealed for the first time after stacking the five observations.
}
\label{fig:field}
\end{figure*}

Using the CIAO tool {\it wavdetect}, we detected 142 point sources across the whole field in the 0.5--2 keV band, supplying the 50$\%$ encircled energy fraction (EEF) PSF map and adopting a detection threshold of $10^{-6}$ and scales of 1, 2, 4, 8 and 16 pixels. A notable population of discrete point sources emerge from the galactic disk including the microquasar and its two hotspots in nebula S26, a well-studied ultraluminous X-ray source (ULX) P13, and a faint point source positionally coincident with the galactic nucleus (see white dashed circles in Figure \ref{fig:field}c).
For the purpose of studying the diffuse X-ray emission, pixels falling in two times the 90\% EER of each detected point source were removed, consistently for the counts, background and exposure maps.

\begin{figure}
\centering
\includegraphics[width=0.45\textwidth]{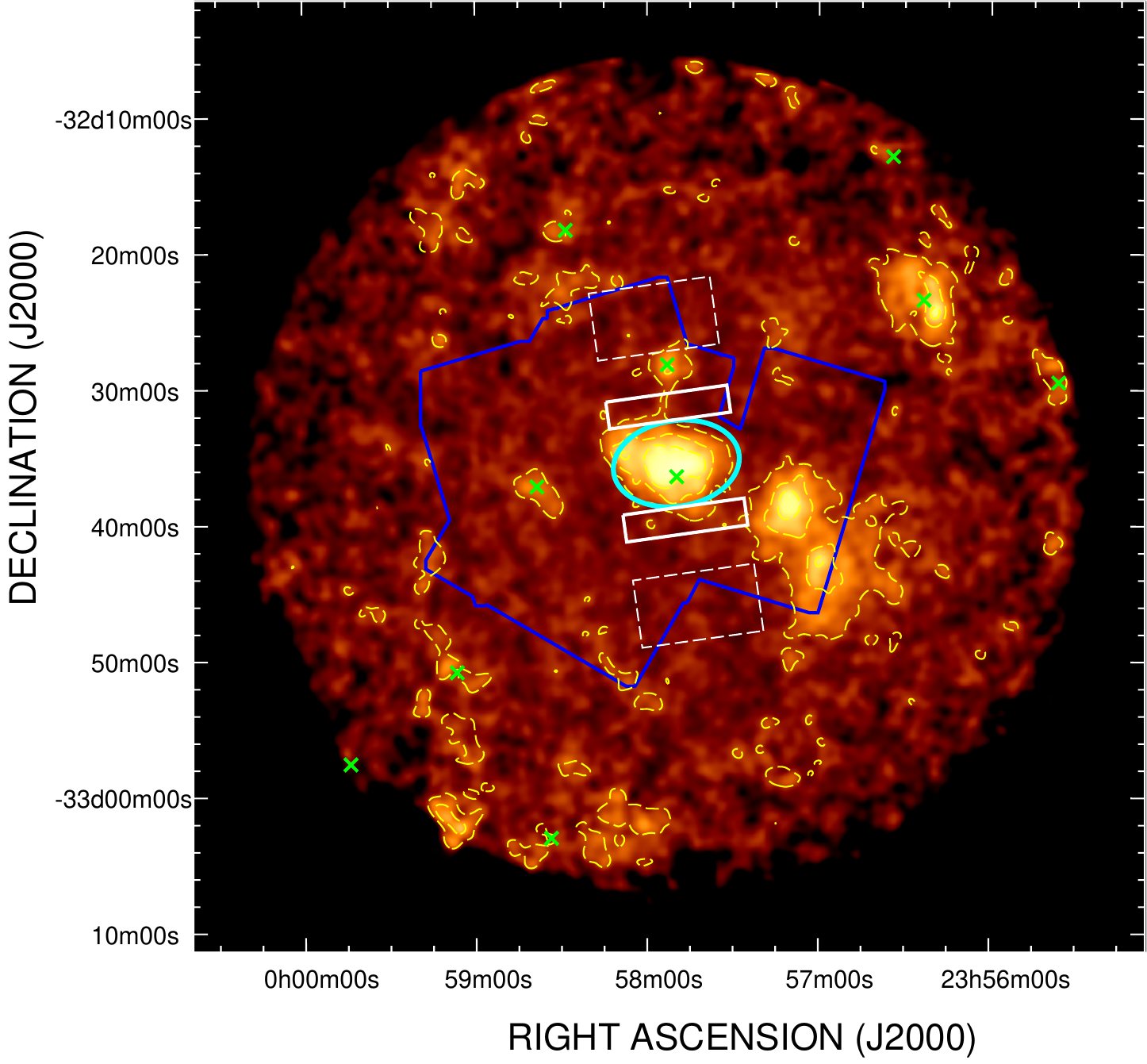}
\caption{eROSITA instrumental background subtracted, exposure corrected image in the energy band of 0.4--2.3 keV. The image has been smoothed with a Gaussian kernel size of 5 pixels, and the point sources have been removed and refilled by sampling their surrounding background. The galactic disk is indicated by the $D_{25}$ ellipse in cyan, while the diffuse X-ray emission is outlined by yellow contours at intensity levels of 3.5, 7.0, 10.5 $\times 10^{-4}\rm~ cts~s^{-1}~arcmin^{-2}$. The two $9\arcmin \times 2\arcmin$ solid rectangles show the spectral extracted regions of the putative north and south halo, while the two dashed rectangles show the adopted background regions. We mark the positions of ten brightest point sources in this field with green crosses. The {\it Chandra} footprint is also shown in blue.
}
\label{fig:fluxmap}
\end{figure}

\begin{figure*}[htbp]\centering
\includegraphics[width=0.49\linewidth]{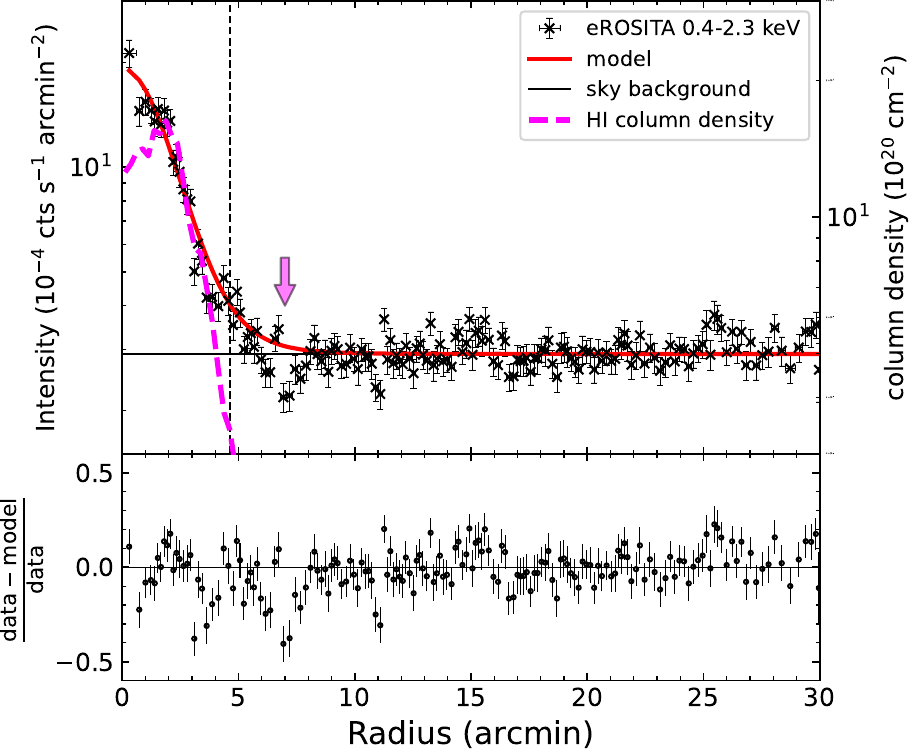}  
\includegraphics[width=0.49\linewidth]{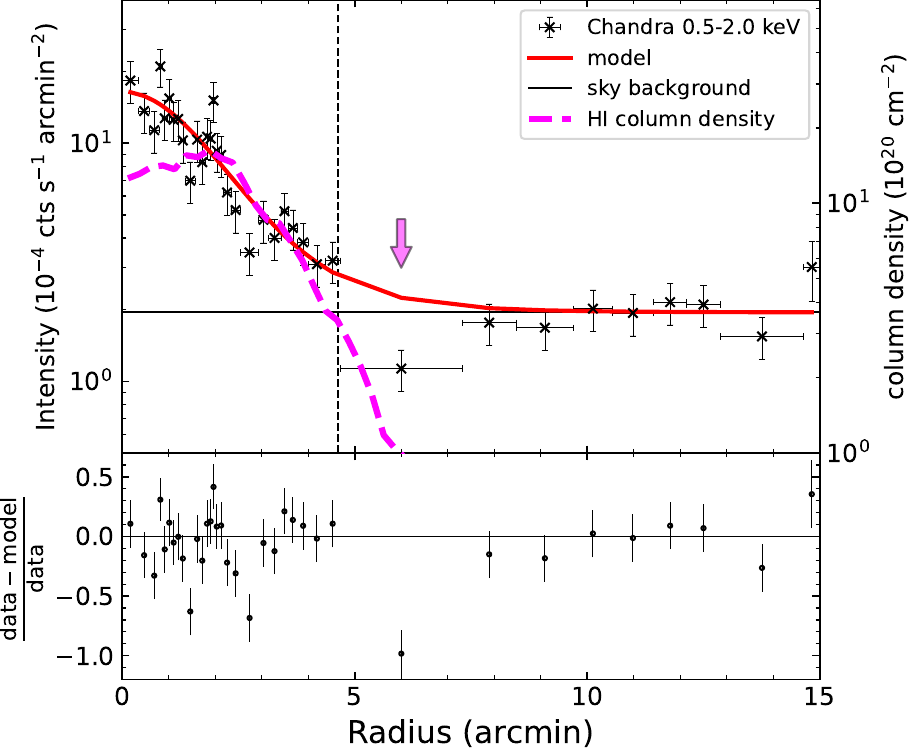}  
\caption{{\it Left}: Instrumental background-subtracted, exposure-corrected radial intensity profile of {\it eROSITA} in the energy band of 0.4--2.3 keV, adaptively binned to achieve a S/N greater than 10. The magenta curve shows the $\textsc{H\,i}$ column density, which is derived from a high resolution VLA $\textsc{H\,i}$ map scaled with a low resolution KAT-7 column density map \citep{Sorgho_2018}. The grey horizontal line indicates the level of sky background. The red curve represents the $\beta$+sky model, which is used to roughly characterize the diffuse X-ray distribution. 
{\it Right}: Same as {\it Left}, but for the radial intensity profile of {\it Chandra} in the energy band of 0.5--2.0 keV, adaptively binned to achieve a S/N greater than 5. 
}
\label{fig:radial}
\end{figure*}

\section{Analysis and results}\label{sec:results}

An eROSITA false-color image of the NGC\,7793 field is shown in Figure \ref{fig:field}a, revealing a large number of point sources as well as extended X-ray features in and around the galaxy, out to a projected distance of 30$\arcmin$ thanks to eROSITA's large FoV. 
In particular, two bright clumps are found to the southwest of NGC\,7793, which have been identified as background galaxy clusters, as label in Figure \ref{fig:field}a. 
A possible third cluster (or galaxy group) is seen near the northwest corner of the field, although we find no available classification or redshift for this feature. 
The {\it Chandra} view of NGC\,7793, which has a smaller field than eROSITA, is shown in Figure \ref{fig:field}c, offering more details on the galactic disk thanks to its superb angular resolution and the deep on-axis exposure. 
We also utilize publicly available data from other wavebands to constrain the distribution of the stellar content, warm ionized gas and atomic hydrogen. 
The composite image shown in Figure \ref{fig:field}b includes $\textsc{H\,i}$ data from VLA (red), $\rm H\alpha$ line emission (green) and B-band data (blue) from the CTIO 1.5-m telescope. We outline the optical disk of NGC\,7793 with a cyan ellipse. The semi-major axis ($a=4.65\arcmin$) and semi-minor axis ($b=3.15\arcmin$) are adopted from the NASA/IPAC Extragalactic Database (NED\footnote{\url{http://ned.ipac.caltech.edu/}}), which is determined from the B-band isophote of 25.0 mag $\rm arcsec^{-2}$ \citep{deVaucouleurs_1991}. The diffuse X-ray emission detected by eROSITA is shown by yellow contours.

\subsection{Radial Intensity Profile}\label{subsec:rprofile}
The eROSITA 0.4--2.3 keV flux image is generated by subtracting the instrumental background from the counts image and dividing the remainder by the exposure map, as displayed in Figure \ref{fig:fluxmap}. For a better visualization, all the removed point sources are refilled by sampling counts from their surrounding background and the image has been smoothed with a Gaussian kernel of 5 pixels in radius. The galactic disk and the brighter part of the diffuse X-ray emission are outlined by contours as in Figure \ref{fig:field}b. 
We mark ten brightest point sources in this field with green crosses. 
Diffuse features are apparent at some of these positions, which might be partly due to the incomplete removal of these bright sources. 
The two or three background clusters also clearly stand out in this image, but they show no clear boundaries, making it difficult to mask out their potential contamination to the putative hot gas halo of NGC\,7793. 
Hence in the following quantitative analysis we discard the western half of the field unless otherwise stated. 

To quantify the extent of the apparently diffuse X-ray emission, we extract the radial intensity profiles for eROSITA in the 0.4--2.3 keV band, as well as for  {\it Chandra} in the 0.5--2 keV band, as shown in Figure \ref{fig:radial}. 
The profiles are adaptively binned to achieve a S/N greater than 5. 
We note that the variation of sensitivity across the FoV has a non-negligible effect on the large-scale X-ray emission. The reduced sensitivity at larger off-axis angles makes it harder to resolve the faint point sources (mostly background AGNs) and hence bring heavier contamination to the diffuse emission, which manifests as an uplift of the radial intensity profile at large radii. 
To correct for this residual cosmic X-ray background (i.e., to achieve a statistically uniform source removal across the FoV), we construct a second background map from the empirical X-ray log$N$-log$S$ relation derived by \citet{Georgakakis_08} (see details in Appendix \ref{sec:appendixA}), taking into account the source detection sensitivity across the FoV. 
The thus created eROSITA radial intensity profile in Figure \ref{fig:radial} has subtracted this residual cosmic background, an effect on the level of 10\%, which is supported by the apparent flatness of the profile at large radii. 
The stacked {\it Chandra} image shows highly non-uniform exposure and hence an even larger variation in the point-source sensitivity across the field. Therefore we have also corrected this effect for the {\it Chandra} intensity profile.

Comparing the two panels of 
Figure \ref{fig:radial}, the eROSITA and {\it Chandra} intensity profiles look qualitatively quite similar to each other, showing an inward rising component corresponding to the galactic emission and a flat component at large radii ($\gtrsim 10'$) representing the local sky background.
On top of this, both profiles exhibit several local dips (or excesses), which can be understood as due to absorption by the spiral arms or the excessive emission from bright point sources. 
However, a noteworthy dip appears at 6$\arcmin$--8$\arcmin$ (indicated by the arrows in Figure \ref{fig:radial}), which is unlikely to be affected by the $\textsc{H\,I}$ disk, as the column density has decreased down to $\rm 10^{20}~cm^{-2}$ at radius of $\sim5\arcmin$ and hence the local absorption cannot explain this dip feature.
Moreover, we have examined the eROSITA profiles in two subbands: 0.4--1 keV and 1--2.3 keV, finding that the dip is even more prominent in the higher energy band, which further argues against being due to absorption. The possibility of instrumental artifact has also been ruled out, as both the eROSITA and Chandra data show the same feature. We discuss the potential origins of this peculiar dip in Section \ref{sec:discussion}.

To have a phenomenological understanding of the diffuse X-ray emission, we fit the radial intensity profiles with a $\beta$-model ($I_X=I_0[1+(r/r_{c})^2]^{0.5-3\beta}$, describing the sum of hot gas and unresolved stellar sources), with an additional constant characterizing the sky background. Both the eROSITA and {\it Chandra} profiles show no highly enhanced central emission and the fitted $r_c$ tends to converge to $\sim 5\arcmin$, suggesting a large core radius of $\sim 5$ kpc.
The best-fit results are $\beta=1.60^{+0.51}_{-0.47}$ and $I_0=16.29^{+0.42}_{-0.42}\times \rm 10^{-4}~cts~s^{-1}~arcmin^{-2}$ for eROSITA, and $\beta=1.18^{+1.69}_{-1.67}$, $I_0=14.55^{+1.58}_{-1.61}\times \rm 10^{-4}~cts~s^{-1}~arcmin^{-2}$ for {\it Chandra}. The level of sky background is indicated by horizontal lines in Figure \ref{fig:radial}. It is shown that the diffuse X-ray emission can be traced out to $5\arcmin$--$6\arcmin$ (equaling to 5--6 kpc in physical scale), which is up to $30\%$ more extended than the starlight, in spite of the existence of the peculiar dip.

\subsection{Vertical Intensity Profile}\label{subsec:vprofile}

To study the X-ray emission genuinely from the so-called hot corona, contributions from the galactic disc, which often overwhelm the former and involve various unresolved stellar sources, have to be taken into account appropriately. For edge-on galaxy, it is straightforward to differentiate the corona from the disc according to the scale height of galactic plane (typically $\sim$ 1 kpc). With an inclination of $50\degree$ \citep{Tully_1988}, NGC\,7793 is close to the most moderately inclined spirals in \citet{Li_13a}, which studied a sample of 53 nearby highly-inclined disc galaxies with $i \gtrsim 60\degree$ using {\it Chandra} observations. We thus extract the vertical intensity profiles of eROSITA in the energy band 0.4--2.3 keV, as well as the {\it Chandra} 0.5--2 keV, $\rm H\alpha$ and $\textsc{H\,i}$ surface brightness profile, to define the vertical range of corona gas. 

\begin{figure*}[htbp]\centering
\includegraphics[width=0.49\linewidth]{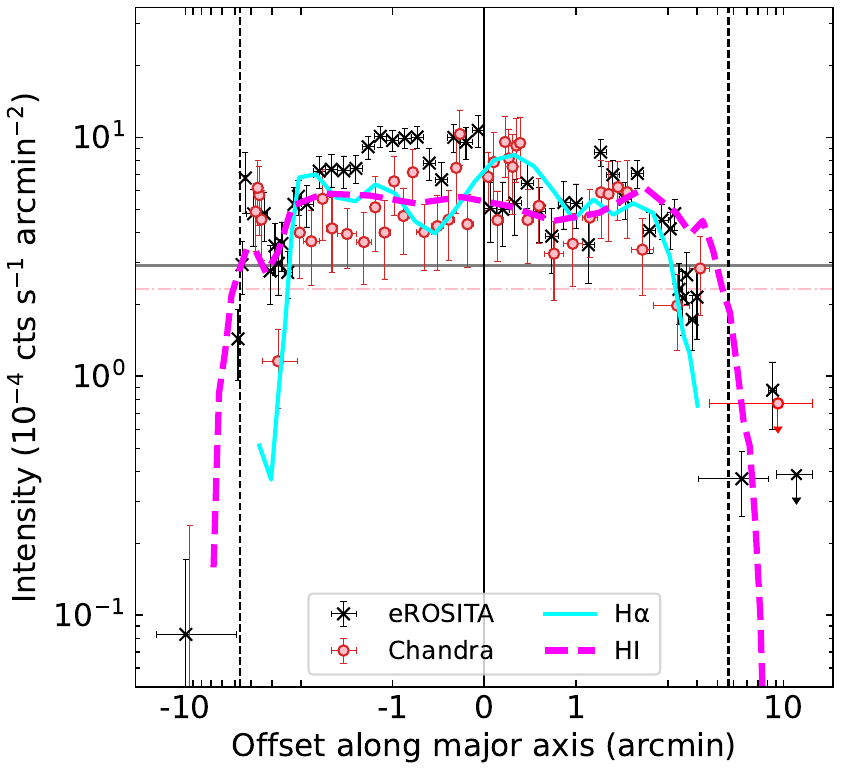}  
\includegraphics[width=0.49\linewidth]{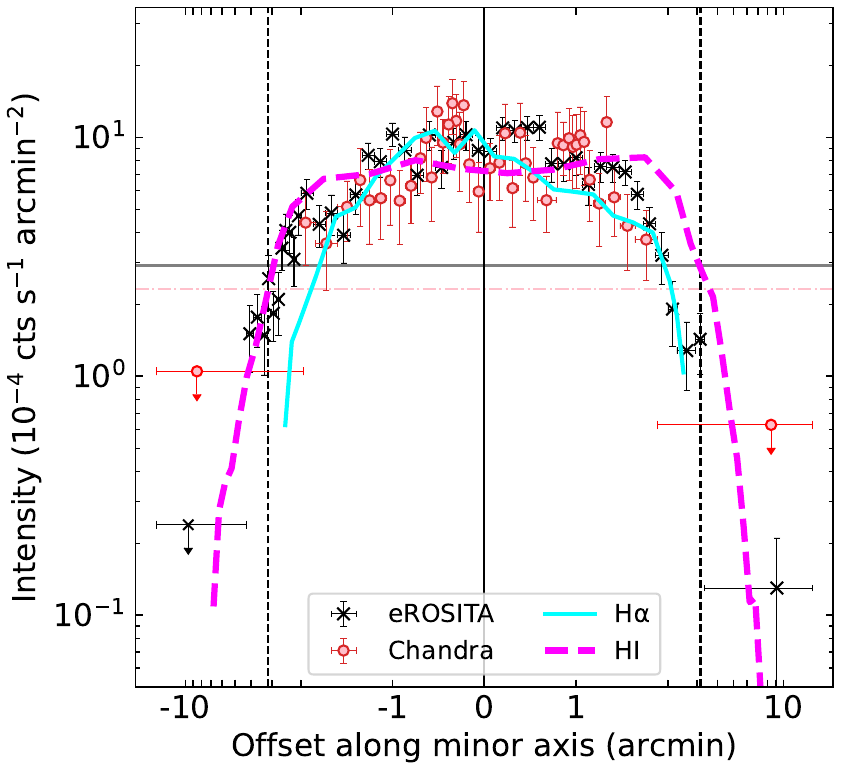}  
\caption{{\it Left}: Sky and instrumental background-subtracted, exposure-corrected vertical intensity profile of {\it eROSITA} along the major-axis in the energy band of 0.4--2.3 keV (black), adaptively binned to achieve an S/N greater than 3. Vertical profiles of {\it Chandra} (red), $\rm H\alpha$ (cyan) and $\textsc{H\,i}$ (magenta) fluxes are overlaid after normalizing to the {\it eROSITA} intensity. The outermost bins, which lack sufficient statistics to satisfy S/N $>$ 3, are represented by upper limits. The horizontal black (pink) line indicates the sky background of eROSITA ({\it Chandra}). The semi-major axis at $\pm4.65\arcmin$ are marked by black dashed lines. We use a `symlog' scale for the x-axis to prevent excessive crowding of data points in the central region, thereby significantly improves the readability.
{\it Right}: Same as {\it Left}, but along the minor-axis. The semi-minor axis at $\pm3.15\arcmin$ are marked by black dashed lines. 
}
\label{fig:vertical}
\end{figure*}

We obtain such profiles along both the major-axis and minor-axis, with a running length of $6\arcmin$ to ensure a nearly full coverage in the {\it Chandra} stacked image. The negative direction corresponds to the east (south) and positive to the west (north), for the major- (minor-) axis. As is illustrated in Figure \ref{fig:vertical}, all the profiles have been renormalized according to the eROSITA intensities within the central $\pm 1\arcmin$ for ease of comparison. It is worth noting that the X-ray data has been subtracted off the sky background to isolate the pure galactic contribution.
The {\it Chandra} profile, although less informative due to the steeply declining exposure at large offset, is consistent with the eROSITA profile. The horizontal lines are identical to those in the radial profiles and indicate the sky background level. 
Notably, the diffuse soft X-ray emission shows some asymmetry in both the central and off-disk regions along different directions, which may arise from residual contamination by point sources and/or from localized feedback processes.
At least a few abrupt excesses are attributed to the the incompletely removed sources: Along the major-axis, the bump at $\sim -4\arcmin$ corresponds to a bright source first discovered by \citet{RP_99} (P9 in that work), whose identity is still uncertain and the X-ray spectrum is very steep ($\Gamma \sim 2.4$) and absorbed. Another bump at $8\arcmin$--$9\arcmin$ on the positive side most likely arises from the northern edge of the background cluster (z=0.2729). 
Nevertheless, the X-rays averagely exceed the scaled $\rm H\alpha$ emission from the inner $3\arcmin$ along both the major- and minor-axis, suggestive of the presence of extraplanar diffuse X-ray emission. The $\textsc{H\,I}$ emission of NGC\,7793 otherwise stretches far, with extension comparable to the X-ray along both directions. This extended $\textsc{H\,I}$ disk of NGC\,7793, as well as its truly declining rotation curve, has been reported by \citet{CP_90}, hinting at a centrally concentrated dark matter halo. 
When incorporating the vertical profiles and the X-ray contours (see Figure \ref{fig:fluxmap}), we note that the extraplanar X-ray emission is not manifested as a bipolar or biconic feature. This suggests that the underlying physical origin of the hot gas may differ from the classical vertical outflows observed in some archetypical superwind galaxies (e.g., NGC 253, M82). However, more dedicated investigations, such as azimuthal analysis, are currently hindered by the relatively low photon counts and the presence of background clusters in the vicinity.

\subsection{Spectral Analysis}\label{subsec:spectra}

Since we are mainly interested in the genuine hot corona of NGC\,7793, contributions from the galactic plane, which dominate the diffuse X-ray emission, need to be marginalized in the spectral analysis. The spectral extraction was performed using the {\it srctool} task. We use two symmetrical rectangle regions with length of $9\arcmin$ along major-axis and inner-to-outer side of $\pm3.15\arcmin$--$5.15\arcmin$ along minor-axis as the source regions (i.e. the north corona and south corona), deliberately avoiding the stellar disk. The corresponding background spectra are extracted from two box regions parallel to the source region, each with a size of $9\arcmin\times 5\arcmin$. The white dashed boxes in Figure \ref{fig:fluxmap} outline the extraction regions, within which all discrete point sources have been carefully removed for the spectral analysis. 

\begin{figure}[htbp]\centering
\includegraphics[width=0.9\linewidth]{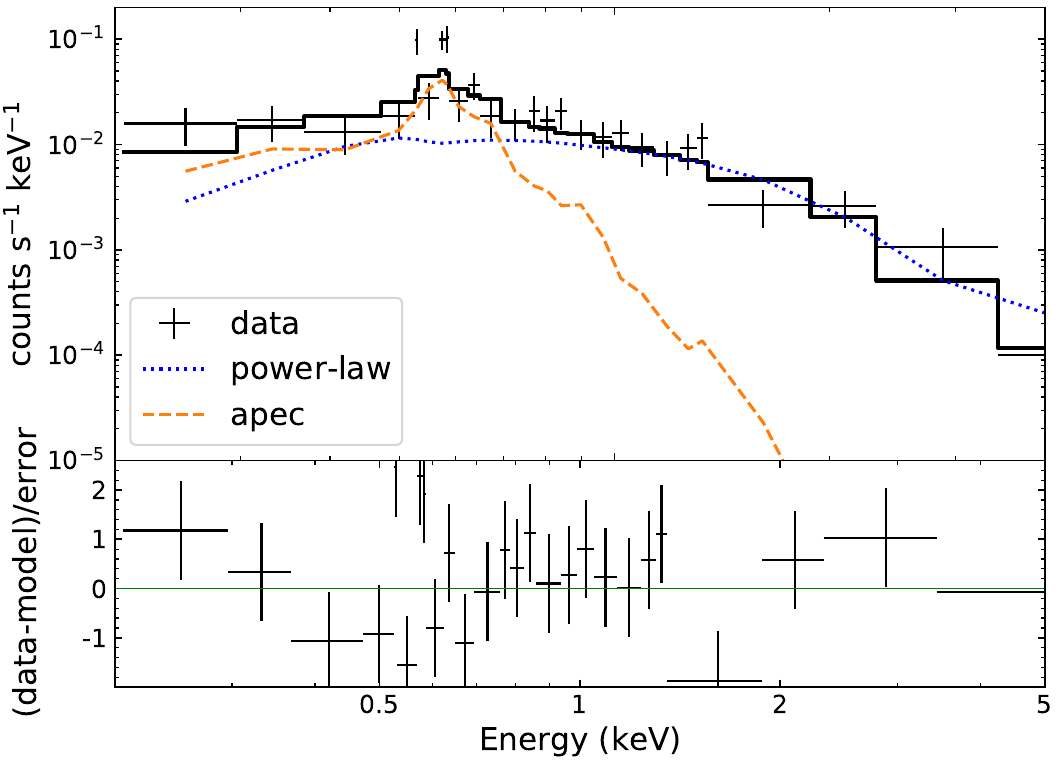}  
\caption{eROSITA spectrum for the extraplanar unresolved X-ray emission of NGC\,7793 and the best-fit model. The spectral data has been regrouped to achieve a signal-to-noise ratio of 3. The fitted model is $\it tbabs\times (powerlaw+apec)$, and the power-law and apec components are shown in blue and orange, respectively. 
}
\label{fig:spec}
\end{figure}

The spectra are adaptively regrouped to achieve a signal-to-noise ratio greater than 3. We adopt an optically thin plasma model (APEC in {\it Xspec}) to characterize the diffuse hot gas, with an additional power-law component needed to describe the emission from the residual point sources (i.e., the XRBs and residual CXB). Both components are subject to the Galactic foreground absorption toward NGC\,7793 ($N_{\rm H} = \rm 3\times 10^{20}~cm^{-2}$). The abundance is poorly constrained and thus fixed at half solar. We find that this model yields similar power-law photon-index and plasma temperature for the spectrum of the north and south corona, with the difference in fluxes being less than a factor of two. Hence the spectra have been co-added to achieve a higher S/N and better constrain the parameters. The combined spectrum is displayed in Figure~\ref{fig:spec}, with the best-fit model characterized by a temperature of $0.18^{+0.02}_{-0.03}$ keV for the hot plasma (orange) and a photon-index of $1.67^{+0.34}_{-0.38}$ for the power-law component (blue). Although the fit is not so satisfactory ($\chi^2/d.o.f=34.22/23$), the data quality is insufficient to introduce more complicated models, thus we adopt it since the results appear to be physical and reasonable.

Surprisingly, the derived halo gas temperature is similar to that of the warm-hot phase gas in the Milky Way’s halo \citep{Yoshino_2009, Gupta_2021, Das_2021, Kaaret_2020, Locatelli_2024, Bhattacharyya_2023}. However, we note that the local background (including the emission from the Galactic halo, the Local Hot Bubble and unresolved CXBs) has been subtracted from the source spectrum, so contamination from the Milky Way's hot gas halo is minimal and the detected 0.18 keV hot plasma is most likely to originate from NGC\,7793 itself. This similarity in temperature may be taken as a hint of the ubiquitous hot halo around $L_*$ galaxies.
We estimate the enclosed hot gaseous mass from the APEC normalization, which gives a value of $\rm \sim 1\times 10^7\ M_{\odot}$. By comparison, the neutral gas mass of NGC\,7793 is $M_{\rm \textsc{H\,i}}\sim 6.8\times 10^8~\rm M_{\odot}$ (\citealt{Walter_08}, a slightly higher value was given by LVHIS, \citealt{Koribalski_18}). The unabsorbed 0.5--2 keV luminosity of the hot gas is $1.27\pm 0.33\times 10^{38}~\rm erg~s^{-1}$. The unresolved point sources contribute a 0.5--2 keV flux of $\rm 1.28\pm 0.23\times 10^{-13}~erg~s^{-1}~cm^{-2}$, translating to luminosity of $2.33\pm 0.42\times 10^{38}~\rm erg~s^{-1}$ at the distance of NGC\,7793.   
We highlight that the extraction region occupies a minor fraction in terms of the generically predicted X-ray halo, which is thought to be quasi-spherical and has a large volume. The actual distribution of the hot gas (e.g. its geometry and the volume filling factor), is still unknowable due to the high sky background and the limiting sensitivity.

\section{Comparison with the TNG50 simulation}\label{sec:TNG50}

The detected extraplanar X-ray emission of NGC\,7793, originating from the putative hot gaseous corona, is not common in normal low-mass (both early- and late-type) galaxies \citep{Bothun_1994, Bomans_2001, Hou_21, Hou_2024}. 
Here we compare our X-ray measurements with the influential TNG50 simulation to shed light on the hot gaseous corona around NGC\,7793-like galaxies.

\subsection{TNG50 Simulation and Sample Selection}
We utilize the data from Illustris-TNG \citep{Nelson_19a, Neslon_19b, Pillepich_19}, a suite of state-of-the-art cosmological simulations carried out using the AREPO moving-mesh code. Specifically, we employ TNG50, which is designed to simulate a large, representative cosmological volume at a numerical resolution approaching the modern zoomed-in simulations of individual galaxies. It evolves $2160^3$ gas cells in a $\rm (51.7~cMpc)^3$ box with best spatial resolution of 74 parsecs and a baryonic mass resolution of $\rm 8.5\times 10^4~M_{\odot}$, which is sufficient to provide a detailed look at the gas structure around galaxies. 

We use the last snapshot ($z$=0) of TNG50 to search for simulated galaxies resembling NGC\,7793. Our selection criteria are as follows: (1) $10^{9.2} < M_*/M_{\odot} < 10^{9.6}$ and $10^{11} < M_{\rm halo}/M_{\odot} < 10^{12}$, to construct a representative sample of galaxies in a suited mass range; (2) $M_{\rm gas}/(M_{\rm gas}+M_*) > 0.1$, where $M_{\rm gas}$ is the mass of cold gas with $T < 10^{4.5}$ K, to exclude gas-poor galaxies; (3) primary flag = 1, to exclude satellite galaxies whose halo gas may be strongly impacted by the central galaxy. After these steps, we identify 460 candidates. To further refine our selection to galaxies with clear disk structures, we use the catalog derived from \citet{Du_2019,Du_2020}, which decomposed the simulated galaxies into well-defined structures (i.e., disk, bulge, and stellar halo). We additionally require our simulated counterparts having ``Spheroids" $<$ 0.5 (total mass fraction for kinematically identified bulge and halo structures) to ensure they are disk-dominated. This leaves us a final sample of 297 galaxies, with SFR ranged from 0.03 to 1.41 $\rm M_{\odot}~yr^{-1}$. To clarify, the involved galaxy properties are defined as follows: the stellar mass ($M_*$) refers to the total mass of star particles which are bound to the subhalo, and SFR refers to the sum of individual SFRs of all gas cells in a certain galaxy.

\subsection{Hot Gas Emission in TNG50 Galaxies}
To investigate the X-ray-emitting hot gas in the simulated galaxies, we consider all gas cells with a temperature higher than $10^{5.5}$ K, which is roughly the virial temperature for galaxies with such mass range. We generate X-ray spectral templates using the PyAtomDB code \citep{Foster_20}, which gives the spectrum of hot plasma in collisional ionization equilibrium at a given temperature. Based on the general understanding that metallicity gradients are present in galaxies, the predicted spectra have taken into account both temperature and metallicity of each gas cell. The total X-ray luminosity in a volume is obtained by integrating the emission of all enclosed hot gas cells in the energy range of 0.5--2 keV. 

\begin{figure*}[htbp]\centering
\includegraphics[width=1.0\linewidth]{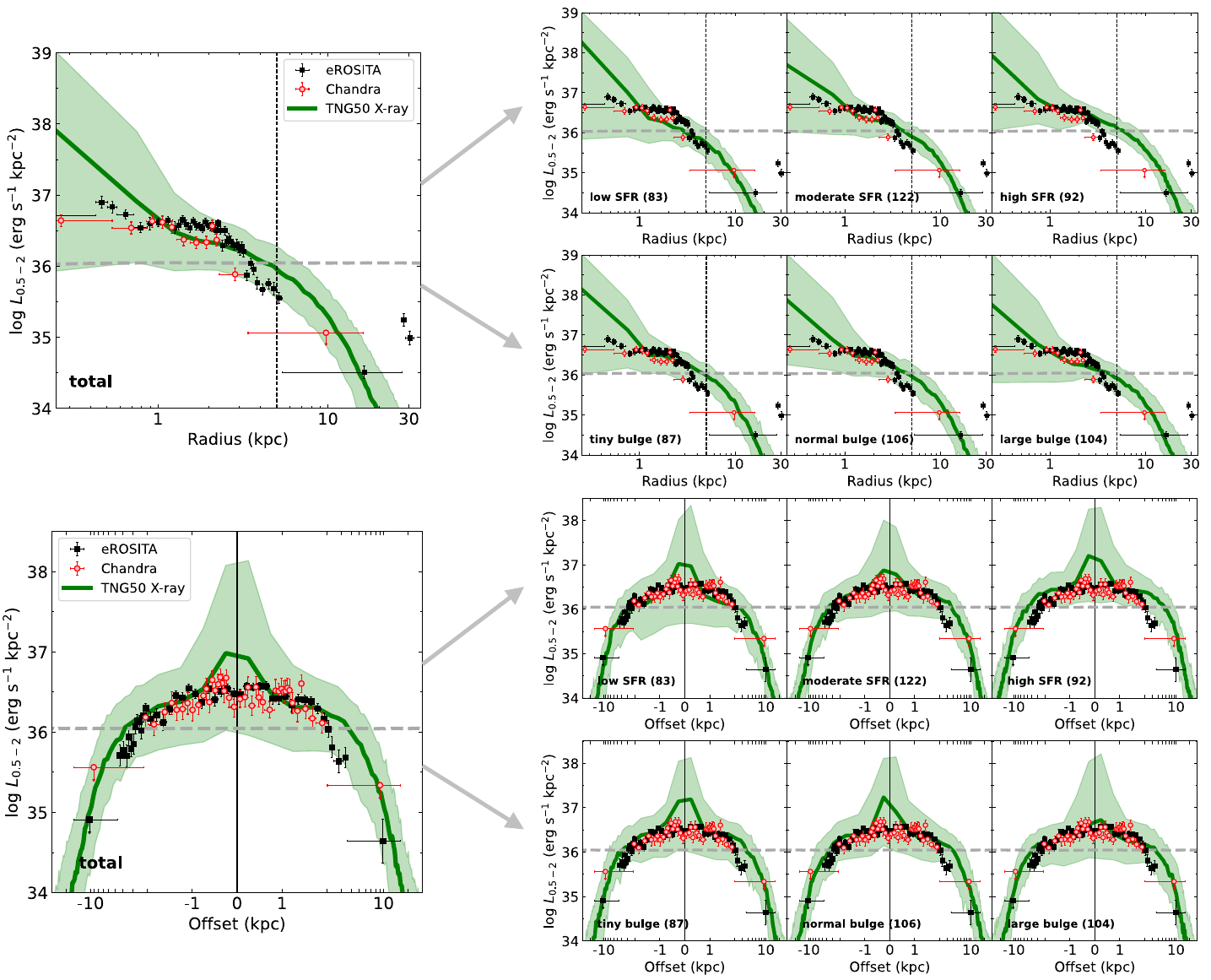}  
\caption{{\it Left}: Radial (Top) and vertical (bottom) X-ray luminosity profiles of the simulated galaxies resembling NGC\,7793 in TNG50, comparing with the observational values (data points) after subtracting sky background. The vertical dashed line marks the $R_{25}$ of NGC\,7793, which is $\sim$ 5 kpc in physical scale. The grey horizontal line indicates the luminosity corresponding to the total background (astrophysical and instrumental background). 
{\it Right}: Same as {\it Left}, but for the subgroups separating by the SFR and bulge mass fraction. The numbers in the brackets indicate the sample size of each subgroup. 
}
\label{fig:TNG}
\end{figure*}

We produce the radial and vertical X-ray luminosity profiles for each simulated galaxy using the particle-level data. To match with the observations, each galaxy was rotated to have an inclination angle of $50\degree$ and projected to the X-Y plane. Given the majority of our sample galaxies are relatively poorly resolved due to their small mass, we adopt a conservative bin size for the radial and vertical profiles, i.e., 0.5 kpc for each bin. A horizontal length of 6 kpc for the vertical bins is adopted to be consistent with the observations.
The predicted profiles from TNG50 simulations are shown in the left column of Figure \ref{fig:TNG}. The overlaid observational data points are obtained by conversing the measured photon fluxes to 0.5--2 keV luminosities, assuming an apec model with temperature and abundance of the best-fit values in Section \ref{subsec:spectra}. We have subtracted the same sky (local) background as in Figure \ref{fig:radial} and Figure \ref{fig:vertical} from the observational data and thus the profiles represent the real emission from hot gas in and around NGC\,7793. 

As is illustrated, the observations are consistent with simulations to some degree, both for the radial and vertical perspective. However, the innermost bins (within 1 kpc) of eROSITA and {\it Chandra} data exhibit a rather flat trend, quite distinct from the simulated results that hot gas emission is notably enhanced in the central region compared to the outer disk. The prediction of brighter-than-observed compact cores in TNG and EAGLE simulations is also reported by \citet{Comparat_22}. From galactocentric radius of 3 kpc, the observational data points start to lie systematically below the median value of simulations, indicating excessive hot gaseous content around low-mass disk galaxies predicted by TNG50 simulations. This discrepancy, however, can also be understood as NGC\,7793 being an exception, given the fact that a representative sample of observed hot gas halo around low-mass disk galaxies is still lacking, which would otherwise yield strong constraints on the galaxy formation models reflected by these state-of-the-art simulations.
Unfortunately, as demonstrated in Figure~\ref{fig:TNG}, the total background (instrumental plus astrophysical background) reaches a value of $\rm 10^{36}~erg~s^{-1}~kpc^{-2}$ at the distance of NGC\,7793, hampering the detection of the predicted hot gas corona at large radii with current data. 

\subsection{Implication of other physical properties}
Besides the low mass, NGC\,7793 is characterized by its high specific star formation rate and the very tiny bulge. We further narrow down the simulated sample by constraining the SFR and bulge fraction (derived from \citealt{Du_2020}), in order to discuss the implication of other galaxy properties on the hot gas emission in TNG50. 

The median SFR ($\rm \overline{SFR}$) and bulge fraction ($\rm \overline{bulge}$) of our sample is 0.36 $\rm M_{\odot}~yr^{-1}$ and 0.18, with standard deviations of 0.2 $\rm M_{\odot}~yr^{-1}$ and 0.08, respectively. We then divided the total sample into low ($\rm \Delta SFR < -0.5\sigma$), moderate ($\rm -0.5\sigma <\Delta SFR < 0.5\sigma$) and high ($\rm \Delta SFR > 0.5\sigma$) SFR subgroups, where $\rm \Delta SFR = SFR-\overline{SFR}$ and $\sigma$ is the standard deviation of SFR. Similarly, we defined bulge subsamples as tiny ($\rm \Delta bulge < -0.5\sigma$), normal ($\rm -0.5\sigma <\Delta bulge < 0.5\sigma$) and large ($\rm \Delta bulge > 0.5\sigma$) based on the bulge mass fraction ($\rm \Delta bulge$=bulge mass fraction $- \rm \overline{bulge}$), with $\sigma$ representing the corresponding standard deviation. Previous studies have suggested that galaxy size determines to a great degree the galaxy evolutionary pathways \citep{Du_2021, Ma_2024} and may also influence the hot gas content. However, the size of NGC\,7793 is not particularly distinctive, and it is less meaningful to construct a representative sample that matches its size. Therefore we restrict our comparison to these two properties and leave a more all-sided discussions for future work.

Results for the subsamples are presented in the right column of Figure \ref{fig:TNG}, in both radial and vertical perspective. 
It is shown that more actively star-forming subgroup tends to have more extended X-ray emission outside the disk, as the deviation from the observations at large radii increasing with higher SFR. For the inner disk emission, the high SFR subgroup aligns better with the observations, though the improvement is slight.
The dependence of hot gas emission on star formation activity in spiral galaxies is predictable, as the Lx-SFR scaling relation has been established and further confirmed in a series of related works \citep{Strickland_04, Li_13b, Li_14}. 
In contrast, we find that the bulge fraction does not have notable effect on the overall level of the hot gas emission. A mild improvement in alignment with the observations is seen in the tiny and normal bulge group. However, the discrepancy between observational and simulated central emission (the innermost 1 kpc) is still significant in all subgroups.
In fact, even the bulgeless galaxies in TNG50 simulation host central BHs more massive than $\rm 3\times 10^6~M_{\odot}$, which is inconsistent with the case of NGC\,7793. This manifestation of the firm lower limit of black hole masses, which makes it difficult to define a sample that exactly matches the observations, is mainly due to the heavy-seed ($\rm \gtrsim 10^5~M_{\odot}$) MBH formation model implemented in IllustrisTNG simulations \citep{Weinberger_18}.

\section{Discussion}\label{sec:discussion}

Based on the newly available data from eROSITA and the archival observations of {\it Chandra}, we have explored the putative hot gaseous halo of NGC\,7793 and found extraplanar diffuse soft X-ray emission on both side of the galactic plane. Comparisons with TNG50 simulations suggest that stellar feedback plays a dominant role in forming and shaping the hot gas in and around low mass spiral galaxies. In theory, the hot gas can originate both externally from the IGM accretion, and/or internally from galactic feedback processes, although it is of significant difficulty to disentangle different components (e.g., the accretion-associated inflows, the feedback-associated outflows and the recycling flows) from the hybrid gas unless high resolution spectroscopy could be achieved. Here we discuss the possibilities of these two origins and examine the prediction on hot gaseous halos from TNG50 simulations and the upcoming X-ray microcalorimeter.

A generic halo of accreted IGM is predicted around present-day massive galaxies, with X-ray luminosity strongly correlated to the gravitational mass and hence scaling with the circular rotation speed of the host galaxy, as $L_{\rm X} \propto V_c^{5-7}$ \citep{Toft_02,Rasmussen_09,Crain_10}. The maximal deprojected rotation velocity of  NGC\,7793 is $\rm \sim 116~km~s^{-1}$ \citep{CP_90}, corresponding to a 0.5--2 keV luminosity of $1-3\times 10^{37}~\rm erg~s^{-1}$, which is one order of magnitude lower than the value we measured. It is not surprising given this galaxy is not so massive as the sample galaxies which \citet{Toft_02} used to derived the $L_{\rm X}$–V$_c$ relationship. Additionally, the hot gas can escape the gravitational potential rather easily if $T > T_{\rm vir}$, inferred to a critical temperature of $\rm \sim 10^5~K$ for this galaxy. Hence we conclude that the IGM accretion plays a minor role for the origin of hot gas in and around NGC\,7793, and thus internal feedback must be taken into account.

It is not clear whether NGC\,7793 harbors a central massive black hole which has impacted its galactic-scale hot gas. On the one hand, nearby star forming dwarf galaxies can indeed host MBHs and the evidences are not rare \citep{Reines_11,Reines_13,Reines_20}. \citet{Pannuti_11} conducted the first high-resolution X-ray study of this galaxy, which utilized a {\it Chandra} ACIS observation (obsID=3954) with exposure of $\sim$ 49 ks and did not detect a central X-ray source associated with a putative MBH. By stacking five {\it Chandra} ACIS observations, we have detected an X-ray nucleus with offset of $0.8\arcsec$ from its optical center. The measured 2--10 keV luminosity is $\rm 1.25^{+0.95}_{-0.82}\times 10^{36}~erg~s^{-1}$, consistent with the estimation in \citet{She_17}, which was derived from the count rate by assuming an absorbed power-law model with column density inferred from the hardness ratio. This value, however, is so low that can also be powered by a normal X-ray binary. Moreover, although a nuclear star cluster (NSC) of NGC\,7793 has been discovered, with a measured dynamical mass of $7.8\times 10^6\ M_{\odot}$ \citep{Walcher_05}, there is currently no robust evidence that a central MBH coexists with the NSC. \citet{Kacharov_18} compared the population $M/L$ ratio of this galaxy to the dynamical estimate and gave a $(M/L)_{\rm dyn}/(M/L)_{\rm pop} \sim 0.5$, which suggests the central MBH will be at the lower mass limit ($10^{4}\ M_{\odot}$) if it does exist. Hence it is highly unlikely that past AGN activities play a significant role in heating and shaping the hot gas in NGC\,7793. 

On the other hand, stellar feedback is the most likely provider of the galactic hot gas in a star-forming galaxy, either through the stellar winds from massive stars or the core-collapse supernovae (CCSNe). Adopting a cooling function $\rm \Lambda=10^{-22.82}~ergs~cm^{-3}~s^{-1}$ for temperature of $\rm T=10^{6.35}~K$ from \citet{Sutherland_1993}, the estimated radiative cooling time for NGC\,7793 is $\sim$ 1 Gyr (the number density of electron is derived from the APEC normalization and has a value of $\sim$ 0.002 $\rm cm^{-3}$), which implies that continuous star formation can significantly contribute to the gas heating and replenishment. Indeed, there are plenty of evidence that NGC\,7793 has undergone recent star-forming activities. 
An HST study covered almost the entire disk has identified 293 young star clusters, among which 65\% is younger than 10 Myr \citep{Grasha_18}. More recently, \citet{Mondal_21} conducted an ultraviolet (UV) imaging study of this galaxy using the Ultra-Violet Imaging Telescope (UVIT) data and found around 61\% of the 2046 far-UV-identified star-forming clumps have age younger than 20 Myr. Thermal energy from massive stars has been found sufficient to explain the large amount of diffuse ionized gas, which is traced by $\rm H\alpha$ emission and often considered to be superposed with the X-ray-emitting hot gas \citep{Strickland_04,Tullmann_06b,Tullmann_06a}. 
From the global perspective, the moderately high SFR (0.52 $\rm M_{\odot}~yr^{-1}$) implies a 0.5--2 keV luminosity of $3-13\times 10^{38}~\rm erg~s^{-1}$ according to the Lx-SFR relation found by \citet{Li_13b}, approximately consistent with the diffuse X-ray luminosity we measured. In addition, feedback from old stellar populations can be more considerable if we assume a Type Ia SNe rate of $0.052[M_*/(10^{10}M_{\odot})][1/(100\ yr)]$ \citep{Mannucci_05}, which amounts to a heating rate of $\sim 1.7\times 10^{40}~\rm erg~s^{-1}$. Although this energy input rate is of two magnitude higher than the measured hot gas luminosity in this work, we note that our estimation has only considered the extraplanar parts of entire diffuse emission (i.e. the genuine hot halo) and hence omit the contribution from ISM gas, which predominates the hot gas deposit and is theoretically hotter and more metal-rich. By chiefly summing the measured hot gas luminosity in this work and \citet{Pannuti_11}, the total estimate can be $\sim 5\times 10^{38}~\rm erg~s^{-1}$, which is still much lower than the expected mechanical energy inputs from Type Ia SNe. This discrepancy can be understood as the shallow gravitational potential being insufficient to confine this hot component that a large volume of gas has been escaped out to the IGM within a rather short time. Unfortunately, given the moderate quality of the halo spectrum, it is impractical to well constrain the abundance, and the fixed sub-solar value (0.5) is likely an underestimate under the scenario of a stellar-feedback-induced metal enrichment. Higher resolution spectra are required to investigate the abundances of different elements (at least the $\alpha$-to-Fe ratio), which helps to assess the respective contributions of Type II and Type Ia SNe in stellar feedback. 

Furthermore, one of our incentives is to place more constraints on the modern numerical simulations by including new measurement of the hot gaseous halo around a less massive spiral galaxy. Based on the robust detection, the hot gas emission displays a rather flat trend within the optical radius in terms of the intensity profiles. The TNG50 simulations, nevertheless, predict a brighter-than-observed core in the innermost region which approximately represents the exponential scale length of the galaxy. This can be interpreted as a sign of AGN-driven outflows in these simulated galaxies, since IllustrisTNG simulations often predict over energetic AGN activities even in the dwarf galaxies (e.g. \citealt{Kristensen_21}). 
From 1 to 3 kpc, where the emission primarily arises from the galactic plane, observations and simulations are in good agreement and further reinforce the key role of the SF-related processes in forming the hot gas. However, at larger radii where the stellar disk is truncated, the observational hot gas luminosity starts to drop faster than the simulated median values. Since our calculations related to the simulated galaxies do not consider the detection threshold, this excessive hot gas emission, even is under luminous at large distance, can always be tracked and actually deviates from real observations. We note that the total background (i.e. the astrophysical and instrumental ones) of eROSITA corresponds to a luminosity of $\rm \sim 10^{36}~\rm erg~s^{-1}$, roughly the value at 4 kpc in the TNG50 profile, making the detection rather challenging as it joins the background and becomes less distinguishable. Nevertheless, the hot gaseous halo around NGC\,7793-like galaxies can be detectable out to radius $\sim$ 10 kpc at a level of $\rm 10^{-4}~cts~s^{-1}~arcmin^{-2}$. By comparison, a study of \citet{Oppenheimer_20} has presented the predictions of hot CGM around normal galaxies using simulated eROSITA stacks in both EAGLE and Illustris-TNG simulations, with results indicating that the hot CGM can be detectable out to 30--50 kpc for intermediate-mass galaxies ($M_* \approx 10^{10.5}~M_{\odot}$) and even 150--200 kpc for high-mass galaxies ($M_* \approx 10^{11}~M_{\odot}$). 
This work, providing the latest detection of hot gas corona around a low-mass disk galaxy, can be considered as a new individual measurement which can be used to compare with the eROSITA stacking measurements, as well as a supplementary evidence for hot corona around galaxy at the low mass end. 

The radial profiles of NGC\,7793 exhibit a dip at $6\arcmin$--$8\arcmin$, which cannot be explained by the absorption of neutral hydrogen and is more prominent in higher energy band. We have attempted to find analogous features in TNG50 simulations, i.e., the local minima in the simulated radial profiles, potentially associated with cavities in the hot gas density and X-ray luminosity maps. This phenomenon can naturally be understood as bubble-like structures produced by past AGN outflows or multiple CCSNe embedded in the galactic disk. However, cosmological simulations help little to verify this scenario due to the relatively poor spatial and time resolution: (i) the abrupt cavities in simulations might just be an effect of the insufficient resolution in these low-mass subhalos; (ii) the time interval between the present (z=0) and last snapshot (z=0.01) is $\sim$ 100 Myr, which makes it impossible to trace how a bubble has been blown and inflated during such a long timescale.  
Although the large-scale X-ray bubbles, perpendicular to the galactic disk, have been observed in the Milky Way \citep{Predehl_20} and predicted to be common-place by numerical simulations (e.g. TNG50, \citealt{Pillepich_21}), we do not find any clear evidence for cavities along the vertical direction of NGC\,7793 in the X-ray flux map. An exceptional bubble structure (S26, \citealt{Pakull_10}) inflated by a powerful microquasar has been discovered in NGC\,7793, yet is much smaller ($\sim$ 300 pc) compared to the width of the dip (a few kpc). Still, two explanations are put forward for this peculiar dip: (1) It originates from a bubble-like structure, where the SNe or AGN-driven winds sweep off the hot gas, creating a cavity less dense than the interstellar medium and the surrounding CGM. (2) It is just a manifestation of the fluctuation of intragroup medium (IGrM), as NGC\,7793 resides in the western end of Sculptor group. 

\begin{figure*}[htbp]\centering
\includegraphics[width=1.0\linewidth]{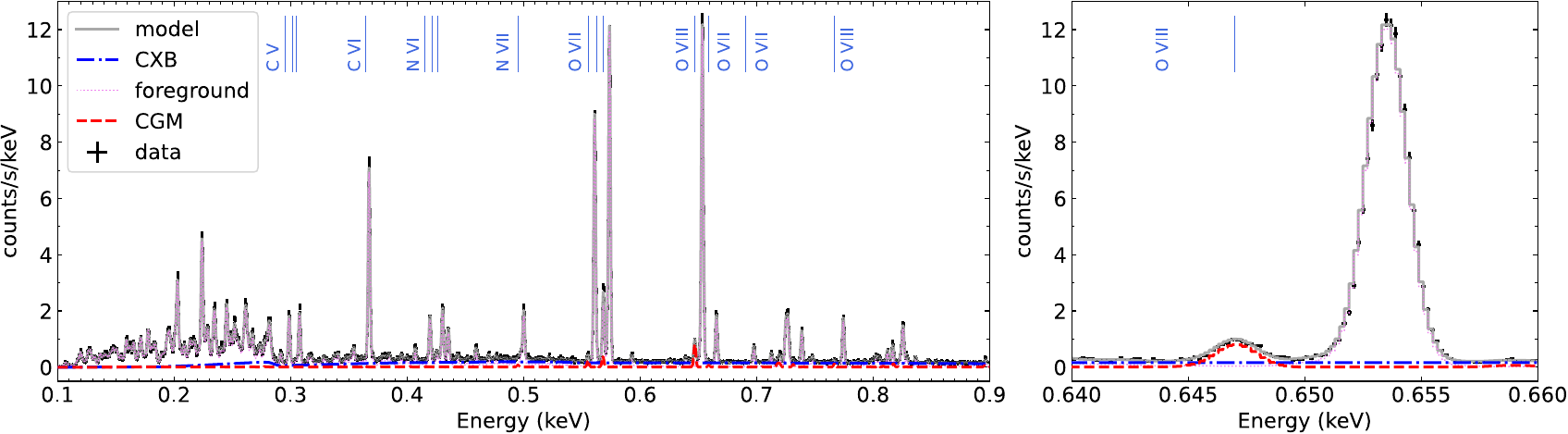}  
\caption{{\it Left}: Mock X-ray spectrum of an example galaxy (subhaloID=605482) in TNG50 (black), which consists of the emission from hot CGM (red), the Galactic foreground (magenta) and the CXB (blue). Emission lines associated with the hot CGM are indicated at the top. {\it Right}: Zoomed-in view of the spectrum around the O VIII $\rm Ly\alpha$ line, which is characterized by the redshifted (assuming $z = 0.01$) line originated from the CGM and distinguishable from the foreground. The primary properties of this simulated galaxy are: (1) $M_*=10^{9.6}~\rm M_{\odot}$; (2) SFR=$1.4~\rm M_{\odot}~yr^{-1}$; (3) $M_{\rm hot}=10^{9.7}~\rm M_{\odot}$.
}
\label{fig:HUBS_spec}
\end{figure*}

Lastly, we examine the capability of the Hot Universe Baryon Surveyor (HUBS, \citealt{Cui_20}) X-ray microcalorimeter to probe the hot CGM around such low-mass galaxies. Designed to have energy resolution of 2 eV and FoV of 1 $\rm deg^2$, HUBS will be an ideal instrument to detect the emission/absorption lines associated with hot CGM and especially suited for studying galaxies at low redshifts. We make mock HUBS observations of NGC\,7793-like galaxies from TNG50 simulations following \citet{Zhang_22}, assuming an exposure time of 1 Ms. Each galaxy has been placed at redshift $\sim$ 0.01 to balance the detection feasibility and the full coverage of its $r_{200}$. The mock X-ray images and spectra are created using the SOXS package\footnote{\url{https://hea-www.cfa.harvard.edu/soxs/index.html}}, taking into account the Galactic foreground and CXB contribution. In brief, the former is modeled as a sum of two thermal models with thermal broadening emission lines ({\it bapec}+{\it tbabs}*{\it bapec}), describing the Local Hot Bubble and the hot halo of our Galaxy. The latter, originating from distant AGNs and galaxies, is modeled by an absorbed power-law. We adopt a uniform radius of $10\arcmin$ for the source regions, which nearly corresponds to the $r_{200}$ for galaxies at z=0.01 and is approximate to the real observations. Pixels within the innermost $0.5\arcmin$ have been removed to exclude emission from the disk. An example of the mock spectrum is displayed in Figure \ref{fig:HUBS_spec}, also shown is the zoomed-in view around the O VIII $\rm Ly\alpha$ line. We use different colors and line styles to differentiate the contribution from hot CGM (red), Galactic foreground (magenta), CXB (blue) and the sum of them (grey). The redshifted lines associated with the CGM are indicated at the top of the figure. It is clear that the emission lines associated with the Galactic foreground are overwhelmingly strong over the whole wavelength range, except for some characteristic lines of the CGM (e.g. the O VIII $\rm Ly\alpha$ line) that are sufficiently significant and can be captured by HUBS. We highlight that the distance of NGC\,7793 is too close to probe its hot CGM, as its $r_{200}$ ($\sim$ 100 kpc, equaling 1.5 degrees in angular) can be far exceeding the FoV of HUBS. In addition, most simulated subhalos in this work have low hot gas luminosity (less than $\rm 10^{38}~erg~s^{-1}$) and hence demonstrate insignificant CGM-associated emission lines in the mock spectra, which are hard to distinguish from the foreground ones if the redshift is only 0.00076 (equals a receding velocity of 274 km/s). 
More specifically, we have found using mock spectra that for NGC\,7793-like halos, the galaxies' O VIII $\rm Ly\alpha$ lines become marginally distinguishable at redshift $\sim$ 0.003, and are fully separated from the foreground lines at redshift $\sim$ 0.01, with the resolution of 2 eV. In the meantime, the idealized targets cannot be too distant, in order to collect sufficient photons to study the properties of the low surface brightness features.
Nevertheless, as is illustrated in Figure \ref{fig:HUBS_spec}, HUBS is a promising probe for X-ray-emitting hot gas around nearby galaxies due to its superior energy resolution and large FoV. Both identification and characterization (e.g., temperature, metallicity, and kinematics) of the hot CGM are within grasp, provided a careful target selection.

\section{SUMMARY}\label{sec:summary}

Based on the early released eROSITA and archival {\it Chandra} observations, we have studied the diffuse soft X-ray emission in and around a nearby low-mass, moderately inclined Sd galaxy, NGC\,7793. Our main conclusions can be summarized as follows:

1. We have detected the extraplanar X-ray emission in energy band 0.4--2.3 keV of eROSITA, and 0.5--2 keV of {\it Chandra}, based on which a generically predicted hot gaseous halo is suggested. Both radial and vertical intensity profiles track the diffuse soft X-ray emission out to $5\arcmin$--$6\arcmin$, or 5--6 kpc from the center, more extended than the stellar disk and the $\rm H\alpha$-traced warm gas. A peculiar dip emerges at $6\arcmin$--$8\arcmin$ in the radial profile, yet the $\textsc{H\,I}$ column density profile suggests it cannot be simply explained by neutral hydrogen absorption. Two scenarios are put forward to explain this dip: (1) it arises from a bubble-like structure; (2) it is a manifestation of the fluctuated IGrM.

2. The eROSITA spectrum of the extraplanar emission can be characterized by an {\it apec}+{\it power-law} model, which gives a best-fit temperature of $0.18^{+0.02}_{-0.03}$ keV for the hot plasma and a photon index of $1.67^{+0.34}_{-0.38}$ for the power-law component. The estimated enclosed hot gaseous mass is $\rm \sim 1\times 10^7~M_{\odot}$, with an unabsorbed 0.5--2 keV luminosity of $\rm 1.3\times 10^{38}~\rm erg~s^{-1}$. 

3. When comparing with the hot gas emission of TNG50 simulated galaxies, observations and simulations are consistent to some degree, further strengthening the detection achieved by eROSITA and {\it Chandra}. However, at large radii (outside the disk), TNG50 simulations always predict excessive emission in our total sample and subsamples. 

4. The estimated hot gas luminosity roughly follows the empirical L$\rm_X$-SFR relation, indicating the major role of stellar feedback in forming and shaping the hot gas halo, wherein the continuous star formation is the most likely provider of the hot gas replenishment. 

5. We examine the capability of HUBS to probe hot CGM around such low-mass galaxies, and find the CGM-associated emission lines are detectable thanks to its 2 eV energy resolution and 1 $\rm deg^2$ field of view. However, galaxies at very low redshifts such as NGC\,7793 may not be ideal targets for HUBS. 

\section*{Acknowledgements}
This paper employs a list of Chandra datasets, obtained by the Chandra X-ray Observatory, contained in the~\dataset[Chandra Data Collection (CDC) 356]{https://doi.org/10.25574/cdc.356}. L.H. would like to thank Amidou Sorgho for kindly providing the HI column density map obtained with the KAT-7 telescope, and thank Teng Liu and Xueying Zheng for helpful discussions about eROSITA data analysis. 
This work is supported by the National Key Research and Development Program of China (No. 2022YFF0503402), the National Natural Science Foundation of China (grant 12225302, grant 12203001), the CNSA program D050102, and the China Manned Space Program through its Space Application System. T.F is supported by the National Natural Science Foundation of China under Nos. 11890692, 12133008, 12221003, and the science research grant from the China Manned Space Project with No. CMS-CSST-2021-A04.

\bibliography{binaryrefs}

\appendix

\section{The CXB correction}\label{sec:appendixA}
We utilize the CXB log$N$-log$S$ relation of the derived from \citet{Georgakakis_08} to correct a residual background in the eROSITA and {\it Chadnra} intensity profiles,  which is manifested as an uplift at large radii and can be attributed to a higher amount of unresolved CXBs at large off-axis angles. The relation follows a broken power-law form as:

\begin{equation}\label{eq:logN-logS}
\frac{dN}{df_X}=
\left\{ 
    \begin{array}{lc}
        K(f_X)^{\beta_1},\ f_X < f_b \\
        K'(f_X)^{\beta_2},\ f_X \geqslant f_b \\
    \end{array}
\right.
\end{equation}

where K and K' are the normalization constants and follows the relation $K'=K(f_X)^{\beta_1-\beta2}$, $f_b$ is the X-ray flux at the break point of the broken power-law, and $\beta1$, $\beta2$ are the power-law indices. Here $f_X$ is in the unit of $10^{-14}~\rm erg~s^{-1}~cm^{-2}$. We adopt the parameter values in the soft band from Table 2 of \citet{Georgakakis_08}, which are $\beta_1=-1.58$, $\beta_2=-2.50$, $\rm log (f_b/10^{-14}~erg~s^{-1}~cm^{-2})=-0.04$ and K=$\rm 1.51\times 10^{16}~deg^{-2}/(erg~s^{-1}~cm^{-2})$.

\end{document}